\documentclass[conference]{IEEEtran}
\IEEEoverridecommandlockouts
\usepackage{cite}
\usepackage{amsmath,amssymb,amsfonts}
\usepackage{algorithmic}
\usepackage{graphicx}
\usepackage{textcomp}
\usepackage{xcolor}
\usepackage[T1]{fontenc} 
\usepackage{fancyhdr}
\def\BibTeX{{\rm B\kern-.05em{\sc i\kern-.025em b}\kern-.08em
    T\kern-.1667em\lower.7ex\hbox{E}\kern-.125emX}}

\begin{document}

\title{Comparing binary and ternary adders and multipliers}

\author{\IEEEauthorblockN{ Daniel Etiemble}
\IEEEauthorblockA{\textit{Computer Science Laboratory (LRI)} \\
\textit{Paris Sud University}\\
Orsay, France \\
de@lri.fr}

}

\maketitle
\begin{abstract}
While many papers have proposed implementations of ternary adders and ternary multipliers, no comparisons have generally been done with the corresponding binary ones. We compare the implementations of binary and ternary adders and multipliers with the same computing capability according to the basic blocks that are 1-bit and 1-trit adders and 1-bit and 1-trit multipliers. Then we compare the complexity of these basic blocks by using the same CNTFET technology to evaluate the overall complexity of N-bit adders and M-trit adders on one side, and NxN bit multipliers and MxM trits multipliers with M = N/IR (IR = log(3)/log(2) is the information ratio). While ternary adders and multipliers have less input and output connections and use less basic building blocks, the complexity of the ternary building blocks is too high and the ternary adders and multipliers cannot compete with the binary ones.
\end{abstract}

\section{Introduction}\label{sec1}

Many papers have been published on the design and performance of ternary circuits at different levels: inverters, basic gates, arithmetic circuits, flip-flops and SRAMs, etc. Some of these papers will be quoted in this text. However, these papers generally do not compare the ternary circuits with the corresponding binary ones. Having presenting such a comparison in 1988 \cite{Etiemble}, we want to do it again more than 30 years later to determine if ternary circuits can compete with the binary ones.
As ternary wires carry more information, N bits are approximately equivalent to M trits according to the relation M = N/IR where IR = log(3)/log(2) = 1.585 is the information ratio per wire.  Table \ref{Trit/bit} presents the correspondence for some values of N. Due to rounding, N/M ranges from 1.5 to 1.6 in  Table \ref{Trit/bit}.

We restrict the comparison to the arithmetic circuits, which are typical implementations of combinational logic. We compare ternary and binary circuits having approximately the same computing capability. For adders or multipliers, it means comparing N-bit circuits with M-trit circuits according to  Table \ref{Trit/bit}.
Adders and multipliers are examined. It turns out that their structure depends on two basic circuits: the 1-bit or 1-trit full adder (FA), and the 1-bit or 1-trit multiplier. 
\begin{itemize}
\item The adders use 1-bit/1-trit full adders
\item The multipliers use 1-bit/1-trit multipliers and 1-bit/1-trit full adders and half adders
\end{itemize}

The rest of the paper is organized as follow:
\begin{itemize}
\item Section 2 compares the number of 1-bit full adders to implement N-bit adders with the number of 1-trit full adders to implement M-trit adders
\item Section 3 compares the number of 1-bit multipliers and 1-bit full adders to implement an N*N bit multiplier with the number of 1-trit multipliers and 1-trit full adders to implement a M*M trit multiplier.
\item Section 4 defines the methodology to compare the complexity of binary and ternary circuits
\item Section 5 compares the hardware complexity of 1-bit full adder and 1-trit full adder and the overall complexity of N-bit and M-trit adders.
\item Section 6 compares the hardware complexity of 1-bit multiplier and 1-trit multiplier and the overall complexity of N x N bit multipliers and M x M trit multipliers.
\item A conclusion summarizes the results of the comparison.
\end{itemize}

\begin{table}
\centering
\caption{Number of ternary and binary wires}
\label{Trit/bit}
\begin{tabular}{|c|c|c|c|c|}
\hline
Number of bits  &8&16&32&64  \\
\hline
 Number of trits  &5&11&21&41  \\
\hline
\end{tabular}
\end{table}

\section{Adders}
\subsection{Carry-Propagate Adders (CPAs)}
\subsubsection{The carry propagate approach}
\begin{figure}[htbp]
\centerline{\includegraphics  [width =8 cm]{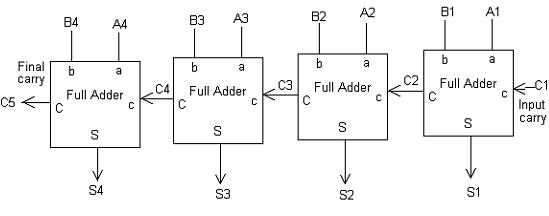}}
\caption{Carry Propagate Adder}
\label{CPA}
\end{figure}

The most straightforward implementation of N-bit or M-trit adders is the carry propagate scheme, also called ``Ripple-Carry Adder". It is presented in Figure \ref{CPA} for a 4-digit adder. For the binary version, Ai, Bi and Ci are binary values. For the ternary version, Ai and Bi are ternary values, while Ci are binary values. In the literature, ternary adders are sometimes presented with ternary carries. We consider them as ternary compressors that could be used in Wallace trees for multiplication. For ternary additions, the carries are always binary.

\subsubsection{8-bit CPA versus 5-trit CPA}
An 8-bit CPA uses eight 1-bit full adder while a 5-trit CPA uses five 1-trit full adder. The comparison is straightforward: a 5-trit multiplier will be more efficient iff the 1-trit adder complexity is no more than x8/5 the 1-bit adder. The result is the same for any M-trit adder compared to a N-bit adder for which $M \approx N/1.585$.
Obviously, there are different other techniques to implement fast adders. Considering all the possible versions is out of the scope of this paper. We just consider two other schemes which purpose is to speed-up the carry propagation. 
\subsection {Carry Look-ahead Adders (CLAs)} 
\begin{figure}[htbp]
\centerline{\includegraphics  [width =8 cm]{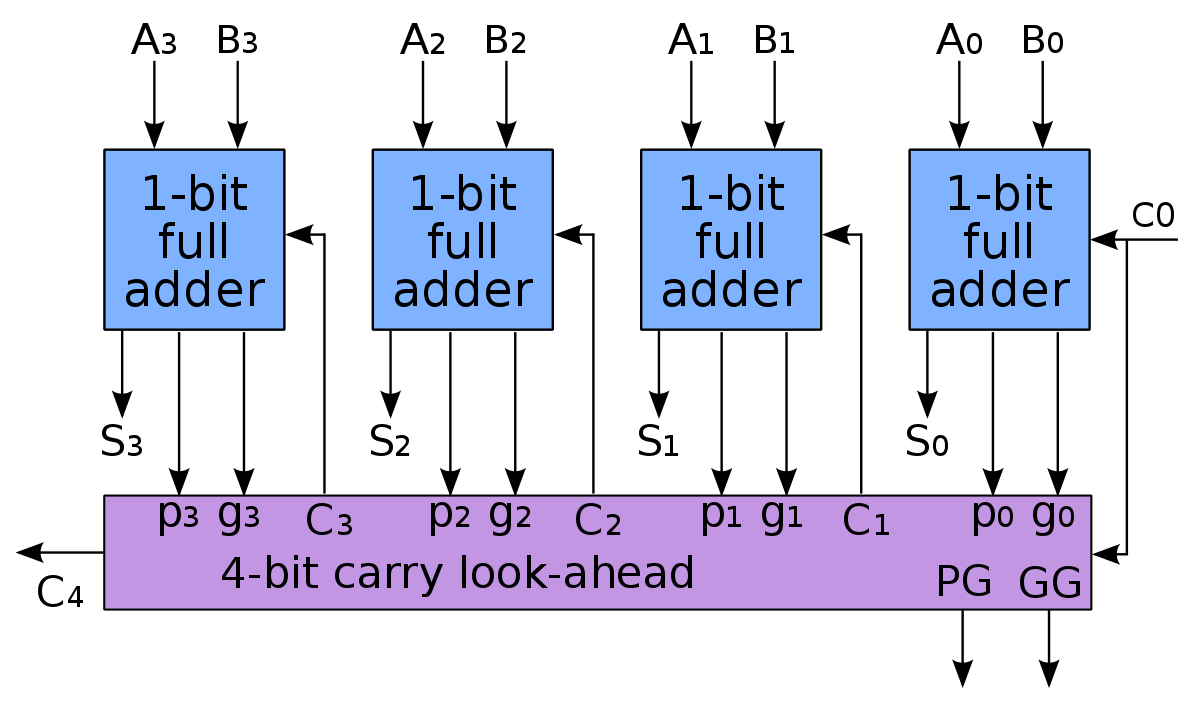}}
\caption{: A 4-bit binary carry look-ahead adder}
\label{4CLA}
\end{figure}

Figure \ref{4CLA} presents a 4-bit carry look-ahead adder. This adder has the same number of full adders than the CPA. The binary equations of the carry computation part are well-known: 
%\begin{equation}
$Gi = Ai.Bi$ \\
$Pi = Ai ~ xor~   Bi$ ~or $Pi=Ai+Bi$\\
$C1= G0 +P0.C0$\\
$C2 = G1 + G0.P1 + P0.P1.C0$\\
$C3 = G2 + G1.P2 + G0.P1.P2 + P0.P1.P2.C0$\\
$C4 = G3 + G2.P3 + G1.P2.P3 + G0.P1.P2.P3 + P0.P1.P2.P3.C0$ \\
%\end{equation}

For the ternary implementation,  Gi and Pi are binary functions of ternary inputs. The computation of C1, C2, C3 and C4 uses binary circuits. 
\begin{itemize} 
\item For stage i, a carry is generated when (Ai = 2 and Bi = 1) or (Ai = 1 and Bi = 2).   
\item For stage i, a carry is propagated when (Ai = 0 and Bi = 2) or (Ai = 1 and Bi = 1) or (Ai = 2 and Bi = 0).
\end{itemize}

Just as CPAs, an 8-bit CLA uses eight 1-bit full adders while a 5-trit CLA uses five 1-trit full adders.
As the binary carry look-ahead computation complexity increases with the carry index, an 8-bit CLA is generally implemented as a cascade of two 4-bit CLAs. The carry computation part is thus two times the carry computation of a 4-bit CLA.
The computation of the carry look-ahead part of a 5-trit CLA can be implemented as a block of 5-trit, with Gi and Pi (0 <i <5) and 5 equations computing C1 to C5.  
Comparing the two approaches means
\begin{itemize}
\item Comparing eight 1-bit full adders with five 1-trit full adders
\item Comparing two blocks of 4-bit look-ahead computation with one block of 5-trit look-ahead computation
\end{itemize}
The comparison is exactly the same for 16-bit, 32-bit or 64-bit adders and the corresponding ternary adders. 

\subsection{Carry Skip Adders (CSA).}

\begin{figure}[htbp]
\centerline{\includegraphics  [width =8 cm]{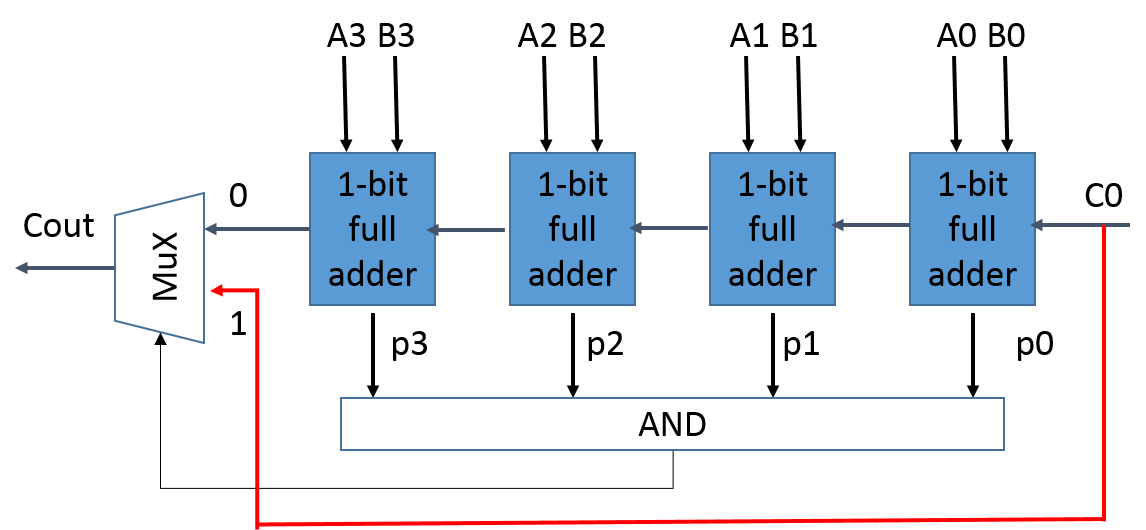}}
\caption{: A 4-bit binary carry skip adder}
\label{4CSA}
\end{figure}

Figure \ref{4CSA} presents a 4-bit binary skip adder. As the CPA, it has four 1-bit full adders. The carry ``skip" scheme uses the Pi propagate function of the CLA, a And gate to compute P0.P1.P2.P3 and a two-input multiplexer. For the ``skip" part, the only difference between the binary and the ternary version is the computation of the propagate functions, which have been defined for the CLA. 

%\subsubsection{8-bit CSAs versus 5-trit CSAs}
Just as CPAs, an 8-bit CSA uses eight 1-bit full adders while a 5-trit CSA uses five 1-trit full adders. As for the CLA approach, the 8-bit ``skip" part is decomposed into two blocks of 4-bit ``skip" parts while the 5-trit CSA has only one block of 5-trit ``skip" part.
Comparing the two approaches means
\begin{itemize}
\item Comparing eight 1-bit full adders with five 1-trit full adders
\item Comparing two blocks of 4-bit ``skip" computation with one block of 5-trit ``skip" computation. A binary block computation uses four Pi functions, one 4-input And gate and one multiplexer. A ternary block computation uses five Pi functions (binary functions of ternary inputs), one binary And gate and one binary multiplexer.
\end{itemize}

\subsection{Adder comparisons}
CPAs, CLAs and CSA have the same N/M ratio of 1-bit adders and 1-trit adders. This ratio is close to IR = 1.585. For CLAs and CSAs, the circuits to speed-up computation slightly modifies the overall comparison. More details will be given in Section 5. 

\section{Multipliers}
\subsection{Comparing a 8*8 binary multiplier with a 5*5 ternary one}
\subsubsection{8*8 binary multiplier}
Typical binary multipliers  can be decomposed in two parts: the first one generates the partial products and the second part reduces the partial products into two sums to be added by a final fast adder. Figure 4 shows the process for an 8*8 bit multiplier.
\begin{itemize}
\item The first part generates 8 partial products of 8 bits, i.e. 64 bits that are the products of $Ai\times Bj$ for $ 0\leq i<8$ and $0\leq j<8$. 
The binary product $Ai \times Bj$ is implemented by a And gate.
\item The reduction of the partials products can use different schemes. The typical one is the Wallace tree, for which 3 lines of partial products are reduced to two lines by using 3-input 2-output full adders (and half adders). For 8*8 bit multipliers, there are several steps of parallel additions of lines of partial products: 8 to 6, 6 to 4, 4 to 3, and finally 3 to 2. A that point, a fast adder is used to add the two final lines. Dadda reduction tree is another one \cite{Townsend}. Other reduction operators can be used, such as 4-2 compressors, 7-3 compressors, etc. With the 3-2 reduction scheme, there are 35 FAs and 18 HAs. Considering the final addition, the total is 44 FAs and 19 HAs.
\end{itemize}

\begin{figure}[htbp]
\centerline{\includegraphics  [width =8 cm]{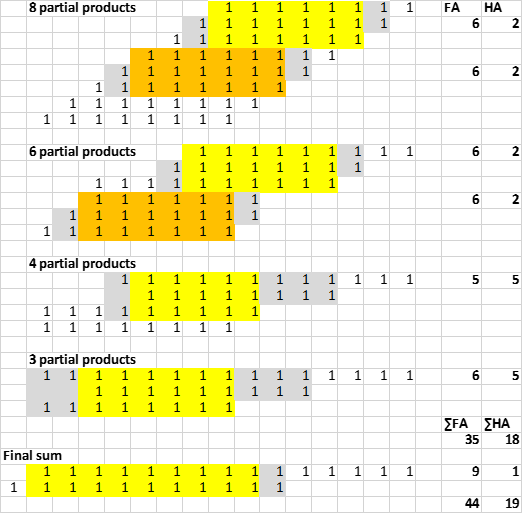}}
\caption{8*8 bit multiplication with a Wallace tree}
\label{88WT}
\end{figure}

\subsubsection {5*5 ternary multipliers}
\begin{figure}[htbp]
\centerline{\includegraphics  [width =6 cm]{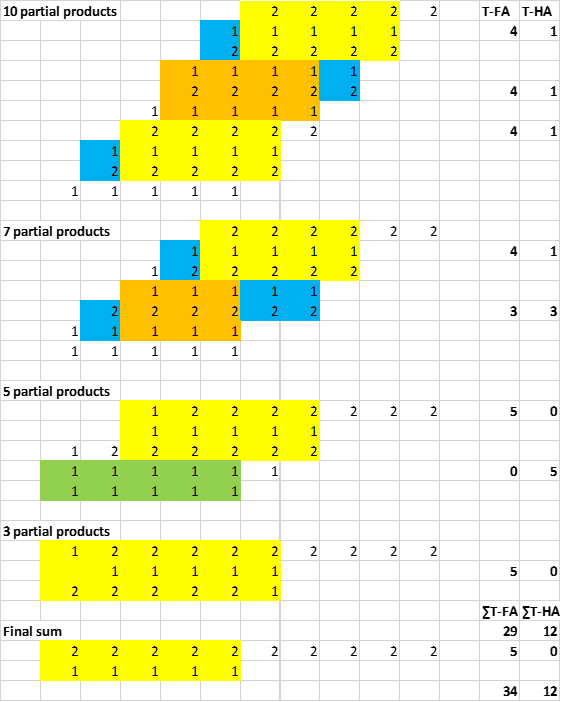}}
\caption{5*5 trit multiplication with a Wallace tree}
\label{55WT}
\end{figure}

Ternary multipliers use 3 different values (0, 1 and 2). A ternary wire carries more information than a binary one. The information ratio is IR = log(3)/log (2) = 1.585. An$ M\times M$ ternary multiplier would be equivalent to a $ N\times N$ binary multiplier with M = N/1.858. For N = 8, M = 5.04. A 5x5 trit multiplier has slightly less computing capability than an 8*8 bit multiplier. 
Figure \ref{55WT} shows the process for a 5*5 trit multiplier.

\begin{itemize}
\item The ternary product $Ai\times Bj$ generates one trit (product) and one binary carry according to Table \ref{1TM}. The first part generates 5 partial products of 5 trits and 5 partial products of 5 bits for $Ai\times Bj$ for $ 0\leq i<5$ and $0\leq j<5$. 10 lines of partial products should be reduced.
\item A Wallace tree can be used to reduce the 10 partial products down to two final lines to be summed by a fast ternary adder. As shown in Figure \ref{55WT}, the reduction process alternates between a set of 2 ternary lines and 1 binary lines (quoted as 2 and 1) and a set of 1 ternary line and 2 binary lines. The reduction tree is thus based on usual ternary full adders (quoted as T-FA) and ternary half adders (quoted as T-HA). There is one specific case when two successive lines are binary lines. In that case, T-HAs can be used for a 2 to 1 reduction that only provides one ternary line. The different steps are thus 10 to 7, 7 to 5, 5 to 3 and 3 to 2 for a total of 29 T-FAs and 12 T-HAs. Considering the final addition, the total is 34 T-FAs and 12 T-HAs.
\end{itemize}

\begin{table}
\centering
\caption{Truth Table of a 1-trit multiplier}
\label{1TM}
\begin{tabular}{|c|c||c|c|}
\hline
Ai&Bi&Pi&Ci\\
\hline
0&0&0&0\\
0&1&0&0\\
0&2&0&0\\
1&0&0&0\\
1&1&1&0\\
1&2&2&0\\
2&0&0&0\\
2&1&2&0\\
2&2&1&1\\
\hline
\end{tabular}
\end{table}

\subsubsection {Comparing N*N bit multipliers with M*M trit multipliers for N =8, 12, 16}
The comparison includes 
\begin{itemize}
\item The number of 1-bit multipliers and 1-trit multipliers
\item	The number of binary full and half adders (B-FA and B-HA) and ternary full and half adders (T-FA and T-HA) for the reduction trees
\item	The final fast adder can use techniques such as carry-look adder adders, carry skip adders, etc. To keep the comparison simple, we consider using the carry propagate adder (CPA) which simply cascades B-FAs or T-FAs. While more sophisticated techniques speed-up the propagation delay, they are still based on B-FAs or T-FAs without changing the number of B-FAs or T-FAs to be used.
\item	Finally, the FA complexity is equal or close to two times the HA complexity. An equivalent number of FAs can be derived assuming that 1 HA = 0.5 FA.
\end{itemize}

Table \ref{C8M}  summarizes the comparison for the $8\times8$ bit multipliers.

\begin{table}
\centering
\caption{Comparison for 8*8 bit and 5*5 trit multipliers}
\label{C8M}
\begin{tabular}{|c|c|c|c|}
\hline
&Binary&Ternary&Binary/Ternary\\
\hline
Ai*Bi&64&25&2.56\\
Reduction FA&35&29&\\
Reduction HA&18&12&\\
Final Add FA&9&5&\\
Final Add HA&1&0&\\
Total FA&44&34&\\
Total HA&19&12&\\
Total equivalant FA&53.5&40&1.34\\
\hline
\end{tabular}
\end{table}

\begin{table}
\centering
\caption{Comparison for 12*12 bit and 8*8 trit multipliers}
\label{C12M}
\begin{tabular}{|c|c|c|c|}
\hline
&Binary&Ternary&Binary/Ternary\\
\hline
Ai*Bi&144&64&2.25\\
Reduction FA&102&102&\\
Reduction HA&34&18&\\
Final Add FA&18&10&\\
Final Add HA&0&0&\\
Total FA&120&112&\\
Total HA&34&18&\\
Total equivalant FA&137&121&1.13\\
\hline
\end{tabular}
\end{table}

\begin{table}
\centering
\caption{Comparison for 16*16 bit and 10*10 trit multipliers}
\label{C16M}
\begin{tabular}{|c|c|c|c|}
\hline
&Binary&Ternary&Binary/Ternary\\
\hline
Ai*Bi&256&100&2.56\\
Reduction FA&200&153&\\
Reduction HA&54&38&\\
Final Add FA&24&13&\\
Final Add HA&1&1&\\
Total FA&224&166&\\
Total HA&55&39&\\
Total equivalant FA&251.5&185.5&1.36\\
\hline
\end{tabular}
\end{table}

Using the same methodology, Table \ref{C12M} compares a $12\times12$ bit multiplier and a $8\times8$ trit one. Table \ref{C16M} compares a $16\times16$ bit multiplier and a $10\times10$ trit one. As 12/1.585 = 7.57, the ternary multiplier has more computing capability. As 16/1.585 = 10.1, the ternary multiplier has slightly less computing capability.

The comparison includes two parts:
\begin{itemize}
\item The number of 1-bit and 1-trit multipliers, which are $N^2$ for N*N bit multipliers and $M^2$ for M*M ternary multipliers. Obviously, $N^2/M^2 \approx IR^2$ = 2.51. Rounding N or M to get integer values for N and M explain the values 2.56 (Table 3 and Table 5) or 2.25 (Table 4). The complexity of 1-trit multiplier versus 1-bit multiplier should not be more that $IR^2$ to get advantage of the ternary approach. 
\item The ternary Wallace tree operates on a smaller number of trits (N/IR), but two times more partial product lines because the 1-trit multiplier generate a product term and a carry. It results than there are only slightly more equivalent 1-bit adders than 1-trit adders. The ratio ranges from 1.13 to 1.36 for the cases that we considered. The complexity of the ternary full adder versus the binary full adder should not be more than a value that is less than IR.
\end{itemize}

\section{Complexity comparisons}
Comparing ternary and binary circuits is not easy. The main reason is that there have been a huge number of binary circuits designed, fabricated and used since the first days of integrated circuits, while very few ternary circuits have been fabricated and used. In the last period, while FinFET technologies have been implemented with 14 nm, 10 nm and even 7 nm technological nodes, only proposals of ternary or multivalued circuits can be found, generally based on simulations. To be able to make significant comparisons, we must define a common technology for both types of circuits and define some complexity measures.
\subsection{A technology: CNTFET}
A carbon nanotube field-effect transistor (CNTFET) refers to a field-effect transistor that uses a single carbon nanotube or an array of carbon nanotubes as the channel material instead of bulk silicon in the traditional MOSFET. The MOSFET-like CNTFETs having p and n types look the most promising ones. This technology has advantages and drawbacks:
\begin{itemize}
\item CNTFET have variable threshold voltages (according to the inverse function of the diameter). Among advantages, high electron mobility, high current density, high transductance can be quoted.
\item	Lifetime issues, reliability issues, difficulties in mass production and production costs are quoted as disadvantages.
\end{itemize}
We use this technology for several reasons:
\begin{itemize}
\item	It is one of the few proposed ones to overcome the limitations of the FinFET technologies after the end of Moore's law.
\item	Its variable threshold voltages make easier the implementation of the different thresholds that are needed for ternary and multi-valued circuits. 
\item	The MOSFET-like CNTFETs have the same circuit styles than the CMOS technologies, which means that the comparison results are not limited to that technology.
\item	A large number of CNTFET ternary or m-valued circuits have been proposed in the recent last years. They facilitate the comparison with the corresponding binary circuits. We will use these proposals for the comparisons.
\end{itemize}

\subsection {Complexity figures}
Hardware complexity is difficult to define as many parameters can be considered:
\begin {itemize}
\item	Number of transistors
\item	Number of interconnections
\item	Chip area
\item	Power dissipation
\item	Propagation delays
\item	Etc.
\end{itemize}
Obviously, the most significant information is speed,  chip area and power dissipation of fabricated chips in a given technology. However, comparing ternary and binary circuits according to chip area and power dissipation is quite impossible as there are very few or no integrated ternary circuits available for comparisons. 
Comparisons must be done with a simple criterion that is available from the circuit electrical scheme. We use the number of transistors. Although the transistor count is only an estimation, it gives significant insights. In fact, when using the same technology to implement the same operator, it is very doubtful that:
\begin{itemize}
\item	More transistors lead to less interconnects as these transistors are interconnected
\item	More transistors lead to less chip area. 
\item	More transistors lead to less power dissipation
\end{itemize}
Finding counter-examples look very challenging!
When the difference in transistor counts is limited to a few \%, no conclusion can be derived. However, if the transistor count for ternary circuits is x2, x3 or more than for the equivalent binary circuits when the information ratio IR = log(3)/log(2) = 1.585, it only means that the ternary circuits have more interconnects, more chip area, more power dissipation than the corresponding binary ones.

\section{Complexity of ternary and binary adders}
\subsection {Complexity of 1-bit and 1-trit adders}
\begin{figure}[htbp]
\centerline{\includegraphics  [width = 8 cm]{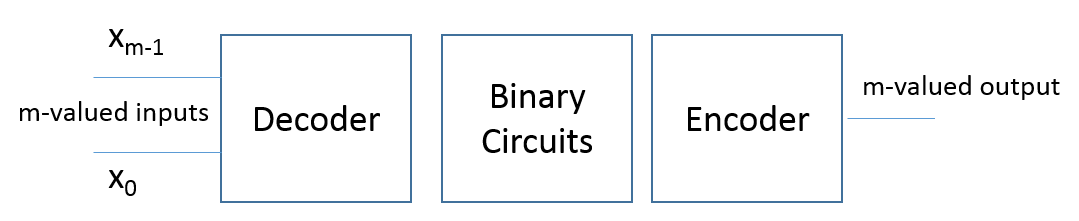}}
\caption{General scheme of m-valued circuits.}
\label{fig}
\end{figure}

\begin{figure}[htbp]
\centerline{\includegraphics  [width = 8 cm]{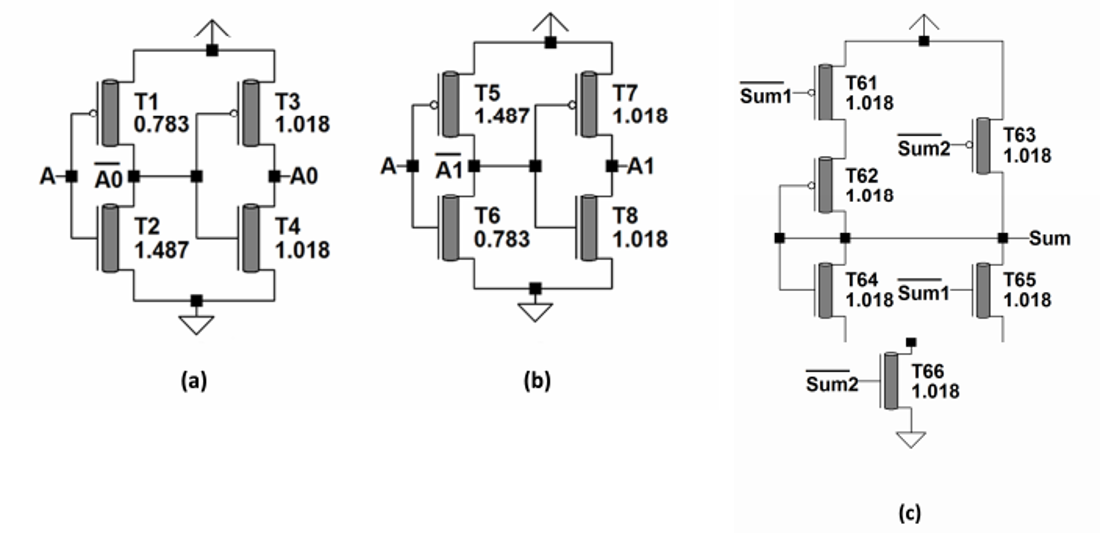}}
\caption{Decoder and encoder circuits \cite{Sahoo}.}
\label{decenc}
\end{figure}

\begin{figure}[htbp]
\centerline{\includegraphics  [width = 8 cm]{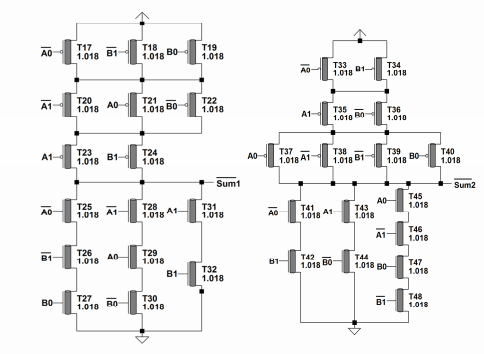}}
\caption{Sum binary parts.}
\label{sum12}
\end{figure}

\begin{figure}[htbp]
\centerline{\includegraphics  [width = 8 cm]{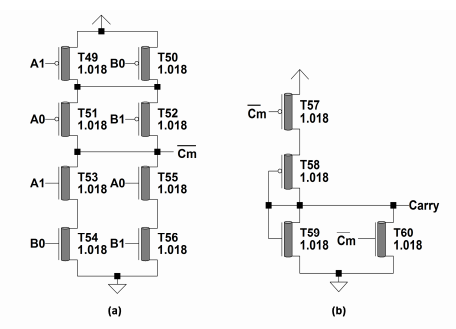}}
\caption{Carry generation.}
\label{carry}
\end{figure}

For 1-bit, various designs have been proposed using binary CMOS circuitry, which can be considered to implement CNTFET binary 1-bit adders \cite{Anitha}. The transistor counts range from 28T for the conventional CMOS design down to 8T for a scheme using 3T Xor gates. Typical implementations with transmission gates use 14T or 16T. All circuits are not equivalent: while conventional CMOS design has maximal noise margins, circuits using transmission gates, or directly connecting inputs either to drain or source of transistors can have reduced noise margins.
There is not a similar comparison of the different ternary 1-trit adders. 
Our reference is the CNTFET ternary half-adder proposed in 2017 \cite{Sahoo} that will be completed to implement a ternary full adder. 
As any multivalued circuits, the ternary half adder uses the general scheme presented in Figure \ref{fig}. The decoder (a, b) and encoder (c) circuits are presented in Figure \ref{decenc}. The sum binary part and the carry generation part are respectively presented in Figure \ref{sum12} and Figure \ref{carry}. The transistor count for the ternary half adder is 66 T.

While the half adder compute Sum10, Sum20 and Cm0 with an implicit input carry equal to 0, similar circuitry can be used to compute Sum11, Sum21 and Cm1 when the input carry is 1. Two multiplexers controlled by the input carry are used to compute the final Sum1, Sum2 and Cm to drive the final encoder and carry circuit. We do not show all the details of computation. The final count for the full adder is 124 T. Table 6 summarizes the number of transistors for the ternary adder and different binary adders. There are from x4.4 to x15.5 more transistors for the ternary adder versus the different binary adders.

\begin{table}
\centering
\caption{Transistor count for ternary and binary adders}
\label{TCHA}
\begin{tabular}{|c|c|c|c|c|}
\hline
&3-FA&Nand-2 FA&Xor-2 FA&8T-FA\\
\hline
Transistor count  & 124  &28&14 to 16&8 \\
Ratio 3/2&  & 4.4  & 7.7 to 8.8 &15.5 \\
\hline
\end{tabular}
\end{table}

\begin{figure}[htbp]
\centerline{\includegraphics  [width = 6 cm]{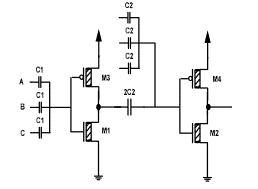}}
\caption{Binary Full Adder with capacitive inputs}
\label{BFadd}
\end{figure}

\begin{figure}[htbp]
\centerline{\includegraphics  [width = 8 cm]{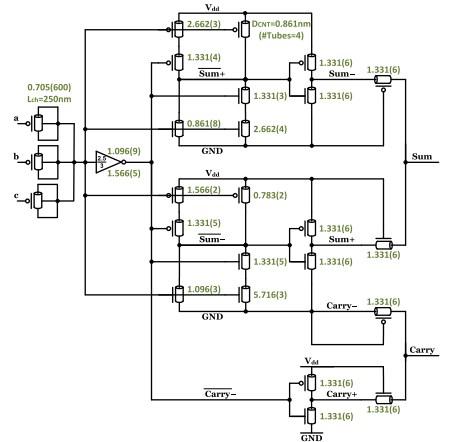}}
\caption{Ternary Full Adder with capacitive inputs}
\label{TFadd}
\end{figure}

Two papers should also be mentioned, which allow a comparison between the binary and the ternary implementations of a carbone nanotube full adder \cite{Navi2}\cite{Navi3}. They both use the threshold logic approach with a linear combination of capacitive inputs. The advantage is the reduction of the number of devices needed to combine the inputs. The drawback is a drastic reduction of the noise margins when coherent noises are simulaneously present on the different inputs. If NM is the noise margin for one input, the noise margin for N-input is NM/N. The binary FA is presented in Figure \ref{BFadd}. Considering that the capacitors are implemented with transistors as in Figure \ref{TFadd}, it has 11 T. The ternary FA is presented in Figure \ref{TFadd}: it has 27T. The ratio is $27/11 = 2.45$, which is greater than the information ratio IR = 1.585.

\subsection{Complexity of carry circuitry}
\subsubsection{Ripple carry adders}
As previously mentioned, a M-trit multiplier will be more efficient than a N-bit multiplier with $M \approx N/1.585$  iff the 1-trit adder complexity is no more than x1.585 the complexity of the 1-bit adder. As shown in Table 6, this is never the case when comparing the best implementations for binary and ternary adders.
\subsubsection{Carry-Look Ahead Adders}
We now compare the carry circuitry for 5-trit and 8-bit CLAs. The equations have been given in 2.2.
Binary Gi and Pi functions are implemented respectively by Nand + Inverter and Nor + inverter. Both function uses 6T. 
Binary C1 is implemented as $C1 = \overline{\overline{G0}.\overline{P0.C0} }$,  i.e. one inverter and two 2-input Nand gates for a total of 10T. C2, C3, C4 are implemented by two levels of Nand gates. The transistor count for a 4-bit carry computation is given in Table \ref{BCCLA}. The transistor count for a 5-trit carry computation is given in Table \ref{TCCLA}.

\begin{table}[!b]
\caption{Transistor count for the carry computations of a 8-bit CLA}
\label{BCCLA}
\begin{tabular}{|c|c|c|c|c|c|c|c|c|}
\hline
Function&Gi&Pi&C1&C2&C3&C4&4-bit&8-bit\\
\hline
Transistor count  & 24 &24&10&18&28&40&144&288 \\

\hline
\end{tabular}
\end{table}

\begin{table}[!b]
\caption{Transistor count for the carry computations of a 5-trit CLA}
\label{TCCLA}
\begin{tabular}{|c|c|c|c|c|c|c|c|c|}
\hline
Function&Gi&Pi&C1&C2&C3&C4&C5&5-trit\\
\hline
Transistor count  & 80 &90&10&18&28&40&54&310 \\

\hline
\end{tabular}
\end{table}

It turns out that the carry computations are more costly for a 5-trit CLA than for an 8-bit CLA. The difference comes from the cost of computing Gi and Pi ternary functions versus the binary ones.

\subsubsection{Carry-Skip Adders}
For an 8-bit CSA, the binary carry computation is composed of two 4-bit skip computations. For 4-bit, it means P0 to P3 functions, a 4-input And gate and a multiplexer. For a 5-trit CSA, the carry computation uses P0 to P4 functions, a 5-input And gate and a multiplexer. The transistor counts are given in Table \ref{CCSA}. Again, the ternary approach is more costly due to the Pi computation costs.

\begin{table}[!b]
\caption{Transistor count for the carry computations of 8-bit and 5-trit CSAs}
\label{CCSA}
\begin{tabular}{|c|c|c|c|c|c|c|c|c|}
\hline
&Pi&Nand+inverter&Mux&4-bit CS&8-bit 5-trit CS\\
\hline
Binary&24i&10&14&48&96\\
\hline
Ternary &90&12&14&&116 \\
\hline
\end{tabular}
\end{table}

\subsubsection {Conclusion for adders}
The transistor count for 1-trit adders is greater than for 1-bit adder and cannot compensate the reduced number of adders. Similarly, the carry computations are more costly for CLAs and CSAs. For CPAs, CSAs and CSAs, the M-trit adders cannot compete with the N-bit adders with M = N/1.585.

\section{Complexity of binary and ternary multipliers}
Ternary and binary multipliers are decomposed in two parts:
\begin{itemize}
\item	1-trit and 1-bit multipliers
\item	1-trit and 1-bit full adders and half-adders that are used in the reduction tree.
\end{itemize}
\subsection{Complexity of 1-bit and 1-trit multipliers}
1-bit multiplier is implemented with a And gate, which means 6T (Nand + Inverter).
1-trit multiplier is far more complicated as it generates a product and a carry according to Table \ref{1TM}. 

Using the same approach as for the ternary adder,
the equations are:
\begin{equation} 
S2= A1.\overline{B1}.B0 + B1.\overline{A1}.A0
\end {equation}
\begin{equation} 
S1= A1.B1 +\overline{B1}.B0.\overline{A1}.A0
\end {equation}
\begin{equation}
Cm=A1.B1
\end{equation}
The number of transistors for the 1-trit multiplier is 4 (decoder) + 12 (Sum2) + 12 (Sum1) + 6 (Product encoder) + 4 (cout encoder) = 38 T, i.e. 38/6 = 6.3. While a 5-trit multiplier uses 25 1-trit multiplier (950 T), a 8-bit multiplier uses 64 And gates for a total of 384 T. The ternary/binary ration is x2.47. There are less 1-trit multipliers than 1-bit ones. However, it cannot compensate their larger complexity compared to the binary ones.

\subsection{Complexity of the reduction tree}
We consider the reduction trees for 5-trit multipliers and 8-bit multipliers. The ternary Wallace tree (Figure \ref{55WT}) has 34 T-FAs and 14 T-HAs and the binary Wallace tree (Figure \ref{88WT}) has 35 B-FAs and 18 B-HAs. For a quick comparison, we can assume that a HA transistor count is half the FA transistor count. Then there are approximately 41 T-FAs and 44 B-FAs. Obviously, this small difference is not able to compensate the advantage of binary FAs compared to ternary FAs that were shown in 2.4. The issue for ternary multipliers is the carry generated by the 1-trit multiplier, which doubles the number of partial products to reduce by the Wallace tree.

\subsection{Overall complexity}
Both the set of elementary multipliers and the Wallace tree needs more transistors for the ternary approach versus the binary ones. M-trit multipliers cannot compete with N-bit ones with M = N/1.585.

\section{Concluding remarks}
We have compared ternary and binary adders and multipliers processing the same amount of information. First, we compared the number of elementary cells such as 1-bit/1-trit full adders, 1-bit/1trit multipliers. This comparison remains valid for any implementation of these cells. Then we consider the hardware complexity using the transistor count for the typical implementation of these cells with CNTFET technology. It turns out that both ternary adders and multipliers cannot compete with the binary ones. If M-trit adders or multipliers have less input and output connections than the corresponding N-bit adders or multipliers, the larger number of transistors means that the ternary arithmetic operators have more connections when considering the internal ones. More transistors mean more connections, more chip area, more propagation delays and more power dissipation for the ternary operators versus the binary ones when using the same technology.

\end{document}